# Ultrathin oxide freestanding membranes with large-scale continuity and structural perfection

Yuhao Hong[1, 2, 3, #, *], Yang Hu[1, 4, #], Jianyao Zheng[1], Minh D. Nguyen[1], Jelle R. H. Ruiters[1], Daniel M. Cunha[1], Iris C. G. van den Bosch[1], Edwin Dollekamp[3], Mark Huijben[1], Yulin Gan[2], Nini Pryds[3], Daesung Park[3], Christoph Baeumer[1], Qinghua Zhang[5], Guus Rijnders[1], Zhaoliang Liao[2, *], Gertjan Koster[1, *]

[1]*MESA+ Institute for Nanotechnology, University of Twente; Enschede, 7500 AE, the Netherlands.*

[2]*National Synchrotron Radiation Laboratory, University of Science and Technology of China; Hefei, 230029, China.*

[3]*Department of Energy Conversion and Storage, Technical University of Denmark, Kgs. Lyngby, 2800, Denmark*

[4]*School of Engineering, Westlake University, Hangzhou, Zhejiang, 310030, China.*

[5]*Beijing National Laboratory for Condensed Matter Physics, Institute of Physics, Chinese Academy of Sciences; Beijing, 100190, China.*

* To whom correspondence should be addressed: yuhho@dtu.dk, zliao@ustc.edu.cn, g.koster@utwente.nl

**Abstract:**

Freestanding oxide membranes offer integration with advanced semiconductor platforms, unlocking opportunities for flexible electronics, silicon-based spintronics, neuromorphic computing, and high-performance energy technologies. Scalable fabrication of such membranes is essential for bridging fundamental discoveries in complex oxides with practical device deployment. However, the lateral dimensions of crack- and wrinkle-free membranes remain limited to millimeter scales, forming a critical bottleneck for large-area architectures. Overcoming this challenge demands strategies that preserve crystalline quality while suppressing defect transfer during release. Here, we demonstrate an approach based on a water-soluble sacrificial layer of super-tetragonal $Sr_4Al_2O_7$, enabling the fabrication of ultrathin, crack-free, and wrinkle-free free-standing oxide membranes spanning centimeter-scale areas. This method is broadly applicable to a wide range of oxides and establishes a new pathway toward large-scale silicon integration and flexible oxide technologies. Nevertheless, dissolution of the sacrificial layer introduces oxygen vacancies into the $SrRuO_3$ membranes, with diffusion depths reaching six unit cells, leading to anomalous "up-and-down" transport behavior. Although post-annealing can eliminate these vacancies, the required temperatures are incompatible with CMOS processes. Therefore, ultrathin freestanding membranes fabricated by water-assisted lift-off still face critical challenges for integration into miniaturized silicon-based oxide devices.

**Introduction:**

Transition-metal oxides, owing to the intricate interplay among their electronic, orbital and lattice degrees of freedom, host a wealth of emergent properties and responses[1], including interfacial superconductivity[2,3], colossal magnetoresistance[4], ferroelectricity[5-7], piezoelectricity[8-10] and a variety of topological textures[11,12]. These functionalities extend well beyond what is achievable in most two-dimensional materials. However, the growth of high-quality single-crystalline oxide thin films typically requires high-temperature epitaxy[13], which is incompatible with advanced semiconductor processing[14]. Moreover, the strong covalent bonding between oxide films and substrates severely restricts their integration with semiconductor platforms[15].

To overcome these constraints, a range of strategies have been explored to prepare freestanding oxide membranes. These include mechanical approaches such as remote epitaxy[16] and controlled exfoliation[17,18], as well as chemical release methods that rely on water-soluble[19] or acid-soluble[20] sacrificial layers. Among them, the water-assisted lift-off method using water-soluble sacrificial layers[19] has emerged as the most practical and cost-effective route. However, many oxides containing Sr and related cations are highly sensitive to water, prone to chemical reactions or slow dissolution[21,22], which inevitably introduce defects into the released membranes. The recently reported super-tetragonal $Sr_4Al_2O_7$ ($SAO_T$) sacrificial layer[23,24] offers a compelling solution: its structural compatibility with perovskite oxides and rapid dissolution substantially shorten the exposure time to water, thereby mitigating defect formation and enabling the fabrication of large-area, crack-free freestanding membranes.

Despite these advances, existing transfer strategies remain challenged by wrinkles and cracks. For example, supporting layers such as polydimethylsiloxane (PDMS) cannot achieve perfect conformal contact, leading to pronounced wrinkling upon strain release [19]. Moreover, mechanical stresses during PDMS removal often induce cracks[25], while organic residues left behind can compromise device integration[26]. Alternative supports, such as spin-coated cellulose acetate butyrate (CAB), afford better adhesion and can be cleanly removed by organic solvents[27], while the high flexibility of CAB makes the membrane insufficiently stabilized during delamination, causing partially detached regions to bend under buoyancy in water and eventually crack. Attempts to reinforce mechanical stability by depositing amorphous $Al_2O_3$ layers[28] can indeed reduce cracking, but at the expense of introducing impurities that hinder device development. Thus, developing a transfer strategy that simultaneously suppresses wrinkles and prevents cracks remains a pressing challenge.

In this letter, we report an innovative approach that leverages the $SAO_T$ sacrificial layer to enable rapid release of oxide membranes while effectively suppressing wrinkle formation. This strategy yields centimeter-scale, crack- and wrinkle-free freestanding regions, and is broadly applicable to a variety of oxides, including ruthenates and ferroelectric titanates. Beyond scalability and structural integrity, our method also permits the fabrication of membranes with an ultimate thickness of ~3 nm, approaching the dimensional limit of advanced Complementary Metal-Oxide-Semiconductor (CMOS) processes[29]. Taken together, these advances establish a robust pathway for the efficient integration of complex oxides with semiconductor technologies.

**Results:**

**Fabrication, transfer and release of freestanding membranes**

$SrRuO_3/SAO_T$ and $BaTiO_3/SAO_T$ heterostructures were epitaxially grown on $TiO_2$-terminated $SrTiO_3$ (STO) (001) substrates by pulsed laser deposition (PLD), with the growth monitored in real time by reflection high-energy electron diffraction (RHEED) (see **Fig. S1**). Distinct RHEED oscillations revealed a layer-by-layer growth mode, while sharp RHEED diffraction spots provided direct evidence for the epitaxial growth of $SAO_T$[24]. As shown in **Fig. 1**, after spin-coating the film surface with the CAB soft supporting layer and covering it with a PDMS stamp, the sample was immersed in deionized water to dissolve the $SAO_T$ sacrificial layer. The $SrRuO_3$ membranes were transferred onto single-crystal silicon substrates, while $BaTiO_3$ membranes were transferred onto Au-coated silicon substrates for ferroelectric characterization. After solvent-assisted removal of the supporting layers, clean, crack-free freestanding membranes were obtained. Details of the transfer process are provided in the Methods section.

**Dissolution kinetics**

**Figure S2** and supplementary videos illustrate the dissolution process of the $SAO_T$ sacrificial layer: deionized water penetrates from the scribed edges and gradually diffuses toward the central region. For SRO sample, the sacrificial layer within a ~5 × 5 $mm^2$ field of view was completely dissolved within ~7 min, accompanied by the formation of wrinkles (see **Fig. S2a-c**). By contrast, in BTO sample the $SAO_T$ layer dissolved within ~6 min under the same conditions, without noticeable wrinkle formation (**Fig. S2d-f**), indicating that the dissolution rate of $SAO_T$ is highly reproducible and essentially independent of the overlying oxide composition. Given that the lattice constant of BTO is larger than those of both SRO and $SAO_T$[23], the results suggest that wrinkling cannot be attributed solely to the rapid release of compressive strain. Furthermore, as shown in **Fig. S2g-h**, no substantial deformation or bending was observed in regions fully covered by PDMS, whereas membranes coated only with CAB exhibited pronounced bending during release, leading to cracks on the edge. This highlights the crucial role of PDMS in preserving the structural quality of freestanding membranes. Finally, the released BTO membrane, adhering to PDMS, retained excellent integrity with a dimension approaching the ~1 × 1 $cm^2$ scale (see **Fig. 2c**).

**Structural perfection**

High-resolution optical microscopy was used to examine the surfaces of freestanding SRO membranes transferred onto Si and BTO membranes transferred onto Au/Si. As shown in **Fig. 2a-b**, the released SRO membranes displayed exceptionally smooth surfaces, with no detectable wrinkles or cracks across central regions. The few dark particles observed are likely introduced during edge scribing and subsequently redeposited after CAB dissolution. Notably, previous studies have reported pronounced wrinkling when using PDMS alone as the supporting layer[23]. Consistent with these findings, we also observed extensive wrinkling and aligned cracking in the central regions of SRO membranes supported only by PDMS (see **Fig. 2f**). By contrast, the introduction of a CAB spin-coated layer effectively suppressed such macroscopic defects, yielding high-quality, wrinkle-free films. Similarly, as shown in Fig. 2d-e, BTO membranes exhibited extremely flat surfaces, showing only a faint scratch near the edge, likely from scribing, while the central areas remained entirely free of wrinkles or cracks. Apart from a few residual particles, no morphological defects were detected, in sharp contrast to earlier reports of severely wrinkled BTO membranes[23]. These results establish CAB spin-coated layer as an efficient strategy to suppress wrinkle formation and markedly enhance the structural quality of freestanding oxide membranes. Meanwhile, the substrate surface remained clean after membrane transfer (see **Fig. S3a**), with no detectable film residues, further confirming the integrity of the freestanding membrane.

Structural characterizations further confirmed the excellent crystalline quality of both the epitaxial films and their freestanding counterparts. Reciprocal space mapping (**Fig. 3a**) showed that the $SAO_T$ and SRO films remain coherently matched with the STO substrate, with no detectable SAO secondary phases, corroborating the high quality of the $SAO_T$ epitaxial films. Moreover, pronounced Laue oscillations observed in the XRD spectra before and after release (**Fig. 3b**) demonstrate that the single-crystalline quality of SRO is well preserved during transfer. Following release, the SRO (002) peak shifts to higher diffraction angles, indicating relaxation of in-plane compressive strain and a reduction of the out-of-plane lattice constant. No $SAO_T$-related reflections are detected, confirming complete removal of the sacrificial layer. The freestanding BTO membranes exhibit a similar strain relaxation, with the (002) peak shifting to higher angles (**Fig. S4a**) and no detectable $SAO_T$ residues. In contrast to SRO, however, BTO does not display clear Laue oscillations. Meanwhile, the thickness of the SRO freestanding membrane, obtained from X-ray reflectivity fitting, is consistent with AFM measurements (**Fig. 3c-e**), confirming the overall thickness uniformity and smooth interfaces. In contrast, the thickness of the BTO freestanding membrane was directly determined by AFM to be ~40 nm (**Fig. S4b**).

**Discussion:**

To meet the dimensional requirements of advanced fin field-effect transistor (FinFET) technologies and next- generation channel devices, the thickness of freestanding membranes must be reduced below 5 nm[30]. However, most reported oxide membranes remain tens of nanometers thick. Although monolayer ferroelectric membranes have recently been demonstrated, their lateral dimensions are limited to the micrometer scale[31]. To explore the thickness limit at the centimeter scale, we directly examined the structural quality of ultrathin freestanding membranes using HAADF-STEM. Remarkably, as shown in **Fig. 4a** & **Fig. S5**, SRO membranes with thicknesses of 8 unit cells (~3.2 nm) and 12 unit cells (~4.7 nm) retained excellent crystallinity, with atomically sharp interfaces to the underlying amorphous $SiO_2$. These observations confirm both the structural integrity of the membranes at the few-nanometer limit and their intimate contact with the Si substrate, without evidence of microscopic wrinkling.

Although we have demonstrated centimeter-scale freestanding membranes that are both ultrathin and structurally coherent, assessing their intrinsic physical properties is essential for evaluating practical applicability. We therefore employed the van der Pauw method to probe the electronic transport of SRO freestanding membranes. As shown in the **Fig. S3b**, macroscopic cracks appeared in the upper-left region of the membrane, likely induced during the transfer process. However, the crack-free lower region extended laterally one centimeter, enabling reliable measurements. As shown in **Fig. 4b**, the epitaxial SRO film exhibits a conventional metallic resistivity-temperature (R-T) behavior with a Curie temperature of ~125 K. In contrast, after release and transfer onto Si, the R-T curve displays an anomalous "up-and-down" behavior between 125-240 K, which is likely associated with certain defects within the freestanding membrane. A similar behavior has previously been reported in SRO channels patterned by Ar-ion milling[32], a process that is known to introduce oxygen vacancies and related defects[33]. Upon annealing at 650 °C, the R-T curve recovered a conventional form, though with a fourfold increase in magnitude and an upward shift of the Curie temperature to ~150 K. Moreover, the negative magnetoresistance was markedly enhanced in the 10-50 K range (see **Fig. S6**). These observations suggest that defects are introduced during the release process but can be partially healed by high-temperature annealing, thereby strongly influencing the transport and magnetic properties of the freestanding membranes.

Focusing on room temperature, we note that the weak resistive switching observed before and after annealing closely resembles the behavior reported in ionic-liquid-gated SRO, where resistivity changes arise from the reversible insertion and extraction of oxygen ions[34]. Consistently, the lattice constant of the released SRO membrane (3.944 Å) (see **Fig. 3b**) exceeds that of bulk SRO (3.93 Å)[35], indicating lattice expansion driven by oxygen vacancies. To directly verify this, as shown in **Fig. S7**, we analyzed ABF-STEM images of a 12 u.c. freestanding SRO membrane prior to annealing. The results reveal that oxygen vacancies are predominantly localized near the $SAO_T$ sacrificial-layer interface, with their concentration increasing towards the $SAO_T$ side, while the CAB spin-coated layer and PDMS-covered surface remains largely vacancy-free. This spatial asymmetry suggests that the vacancies are introduced during dissolution of the $SAO_T$ sacrificial layer, rather than during removal of the CAB spin-coated layer. Consequently, only a small amount of oxygen vacancies are present in as-grown SRO films, but additional vacancies are incorporated upon release, leading to a reduction in resistivity. The vacancy penetration depth is limited to approximately 6 u.c., making them non-negligible in ultrathin membranes. By contrast, in previously reported SRO freestanding membranes with thicknesses of tens of nanometers, such effects are likely averaged out, obscuring the anomalous R-T behavior.

## Conclusion and Outlook

In summary, we report an advanced water-assisted transfer methodology that enables the fabrication of freestanding oxide membranes spanning nearly 1 cm in lateral dimensions, with thicknesses down to the ultrathin limit of 3.2 nm, while remaining free from macroscopic cracks and wrinkles. This method not only suppresses large-scale wrinkling but also demonstrates wide applicability across multiple oxide systems, including ferromagnetic ruthenates and ferroelectric titanates, thereby providing a practical and versatile route for the construction of miniaturized oxide devices directly compatible with silicon platforms.

Nevertheless, dissolution of the $SAO_T$ sacrificial layer introduces a high density of oxygen vacancies in SRO, with diffusion depths up to 6 u.c., giving rise to pronounced anomalous transport behaviors in ultrathin membranes. While post-annealing can mitigate these defects, the required thermal conditions are incompatible with CMOS processing, posing a critical challenge for integrating ultrathin freestanding oxides into silicon-based device platforms. Overcoming this challenge will likely require the development of alternative release strategies that minimize defect incorporation, or the design of oxygen-stable functional oxides that are intrinsically robust to the dissolution process. Addressing these bottlenecks will be decisive for unlocking the full potential of freestanding oxide membranes in next-generation flexible electronics, neuromorphic computing, spin-based logic, and sustainable energy technologies.

## Method

### Sample growth

SRO/SAOT and BTO/SAOT heterostructures were grown on atomically smooth $TiO_2$-terminated STO (001) substrates using pulsed laser deposition (PLD). During deposition of the $SAO_T$ sacrificial layer, a laser fluence of 1.0 J $cm^{-2}$ was applied, with typical substrate temperatures of 750 ºC and an oxygen partial pressure of $5 \times 10^{-3}$ mbar. For the SRO and BTO layers, the laser fluence, substrate temperature, and oxygen pressure were set to 2.0 J $cm^{-2}$, 650 ºC, 0.1 mbar and 1.0 J $cm^{-2}$, 650 ºC, 0.01 mbar, respectively. The growth of each sublayer was precisely monitored in situ using reflection high-energy electron diffraction (RHEED). Following deposition, the films were slowly cooled to room temperature under 200 mbar of oxygen.

### Transfer and release of freestanding membranes

A cellulose acetate butyrate (CAB) support layer was spin-coated onto the sample surface at 3000 r.p.m. for 60 s with an acceleration of 1000 r.p.m. $s^{-1}$, followed by baking at 150 °C for 10 min to cure the film. The sample edges were scored with a surgical blade to expose the sacrificial layer, allowing water penetration. A polydimethylsiloxane (PDMS) stamp (~0.5 mm thick, $13 \times 10$ mm$^2$) was then attached to the surface to provide mechanical support. The sample was immersed in deionized water until the $SAO_T$ sacrificial layer was fully dissolved, releasing the freestanding membrane. Using tweezers, the PDMS-supported SRO and BTO membranes were transferred onto single-crystal Si substrates and Au-coated Si substrates, respectively, for subsequent ferroelectric characterization. The membranes were dried in air at room temperature for ~10 h, followed by annealing at 100 °C for 10 min to improve adhesion. After cooling, the samples were immersed in ethyl acetate (EA) for ~20 min until the PDMS stamp detached spontaneously. Finally, residual CAB was removed by sequential rinsing in acetone and isopropanol (~5 min each).

## Structural characterization and STEM imaging

The XRD, XRR and RSM was characterized by a Panalytical X-ray Diffractometer. AFM images were performed on the Bruker. HAADF-STEM images were performed on an aberration-corrected FEI Titan Themis G2 operated at 300 kV.

## Transport Measurement

All electrical transport tests were performed using the Van der Pauw method in a PPMS (Quantum Design DynaCool system).

## Optical characterization

All optical micrographs and videos were acquired using a Nikon upright optical microscope.

## Data availability

The data that support the findings of this study are available from the corresponding author upon request.

## Acknowledgments

This work was supported by the China Scholarship Council (No. 202306340089 (Y. H.)), the CAS Project for Young Scientists in Basic Research (YSBR-100 (Z.L.)), the National Key R&D Program of China (No. 2022YFA1403000 (Z.L.)) and the National Natural Science Foundation of China (Nos. 52272095 (Z.L.) and 11974325 (Z.L.)).

## Author Contributions Statement

Y. Hong and Y. Hu contributed equally to this work. Y. Hong directed the project, performed the experiments and wrote the manuscript. Y. Hu designed the experiments and wrote the manuscript. J. Z. provided the PDMS stamp. M. N. provided the silicon substrate. Q. Z. contributed to STEM. J.R. and D. C. contributed data analysis. M. H., Y. G., D. P., C. B., G. R., Z. L., G. K. reviewed the manuscript. All the authors discussed the results and commented on the paper.

**Competing Interest Statement**

The Authors declare no competing interests.

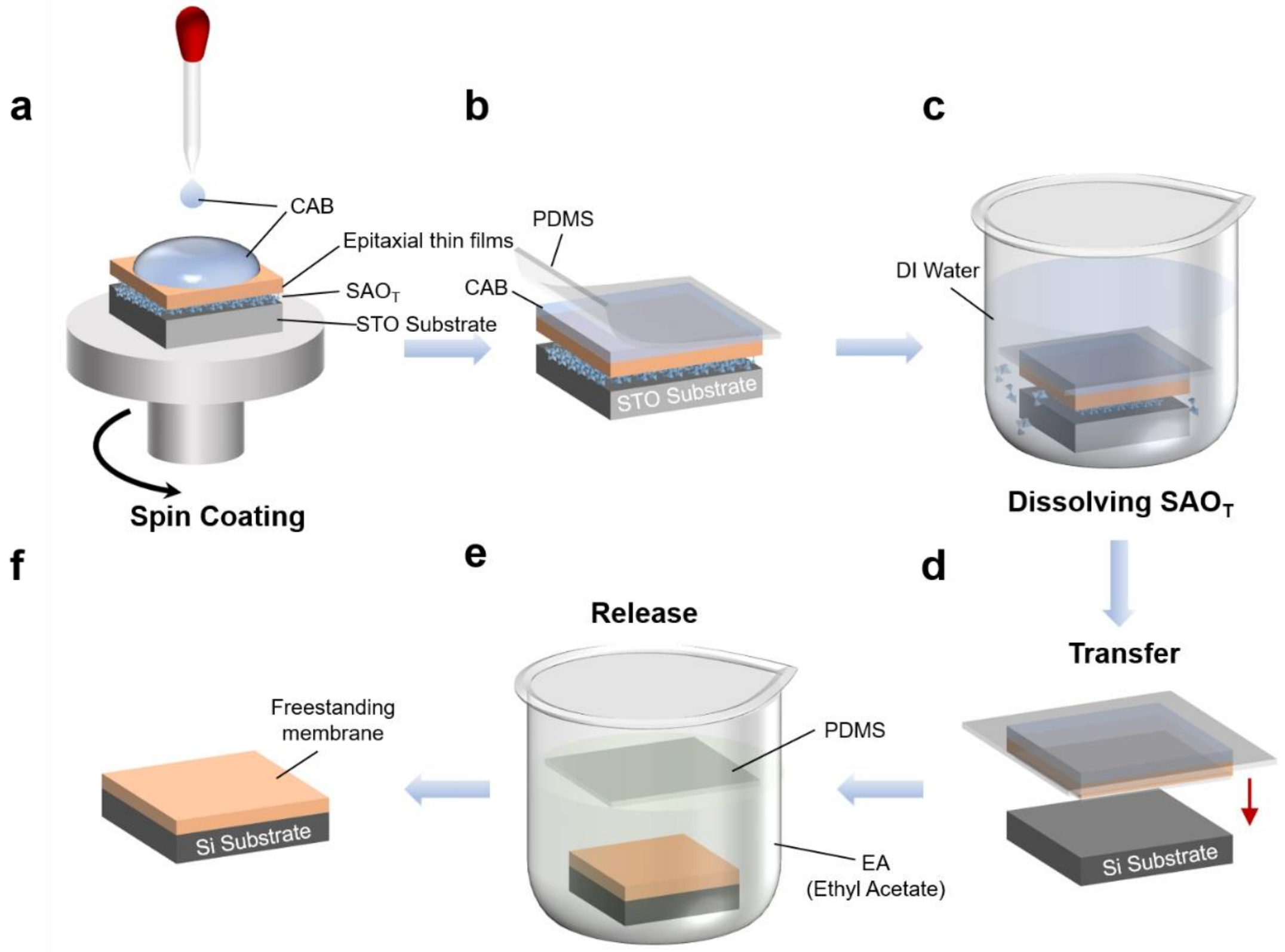

**Figure 1 | Schematic illustration of the transfer and release of freestanding films.** The process begins with **(a)** spin-coating a CAB support layer, followed by **(b)** covering with a PDMS stamp. **(c)** The underlying SAOT sacrificial layer is then dissolved in deionized water, allowing **(d)** the freestanding film to be transferred onto a Si substrate. Finally, **(e)** the CAB layer is removed by dissolution in ethyl acetate, thereby releasing the film, which remains **(f)** seamlessly integrated with the Si substrate.

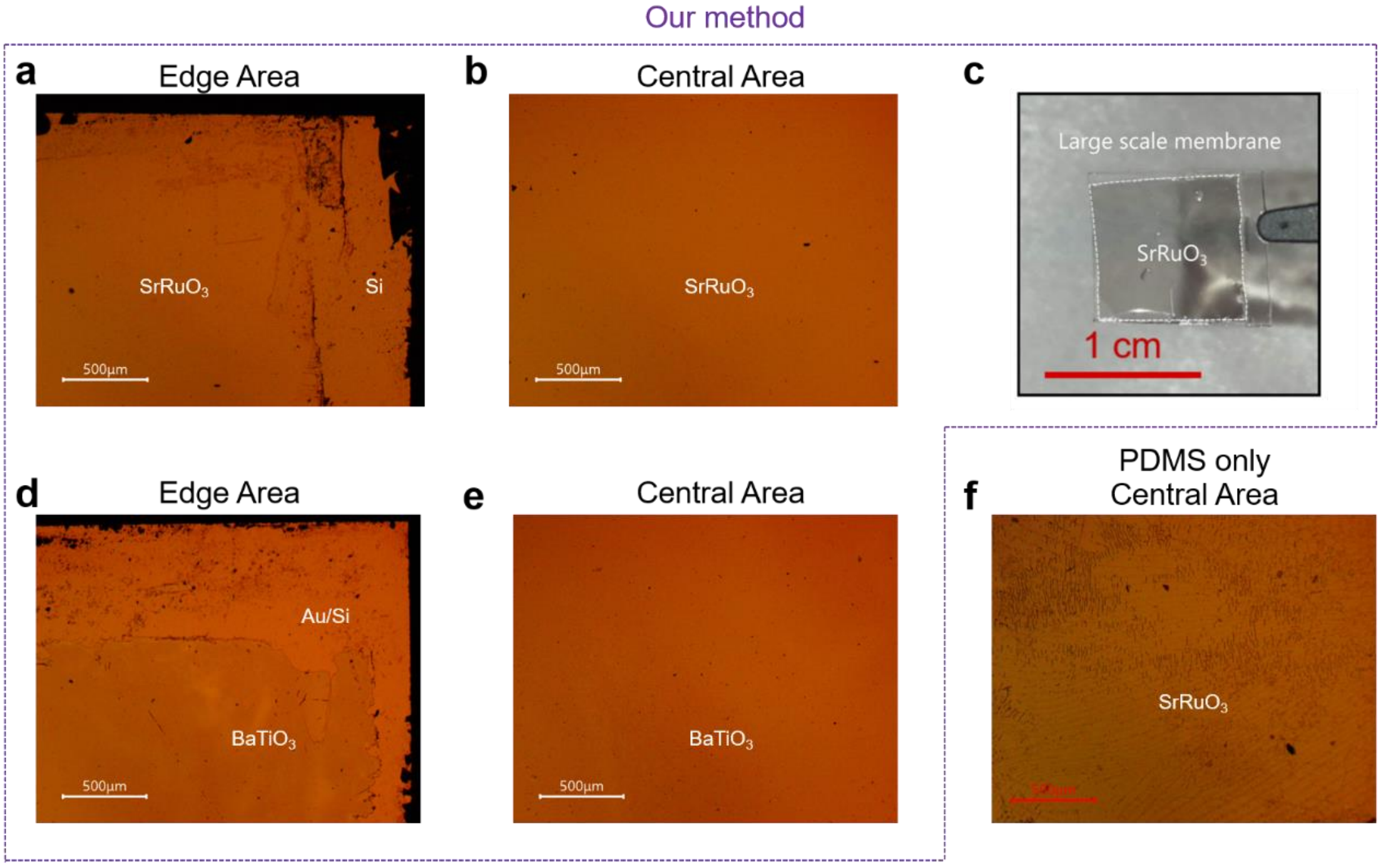

**Figure 2 | Optical morphology of freestanding membranes.** Optical morphology at the **(a)** edge and **(b)** center of an SRO freestanding membrane. **(c)** Optical image of a centimeter-scale SRO freestanding membrane integrated with a PDMS stamp. Optical morphology at the **(d)** edge and **(e)** center of a BTO freestanding film. **(f)** Optical morphology of the center region of an SRO freestanding membrane transferred and released using only a PDMS stamp. Where **a-e** correspond to freestanding membranes transferred and released with both CAB and PDMS as support layers.

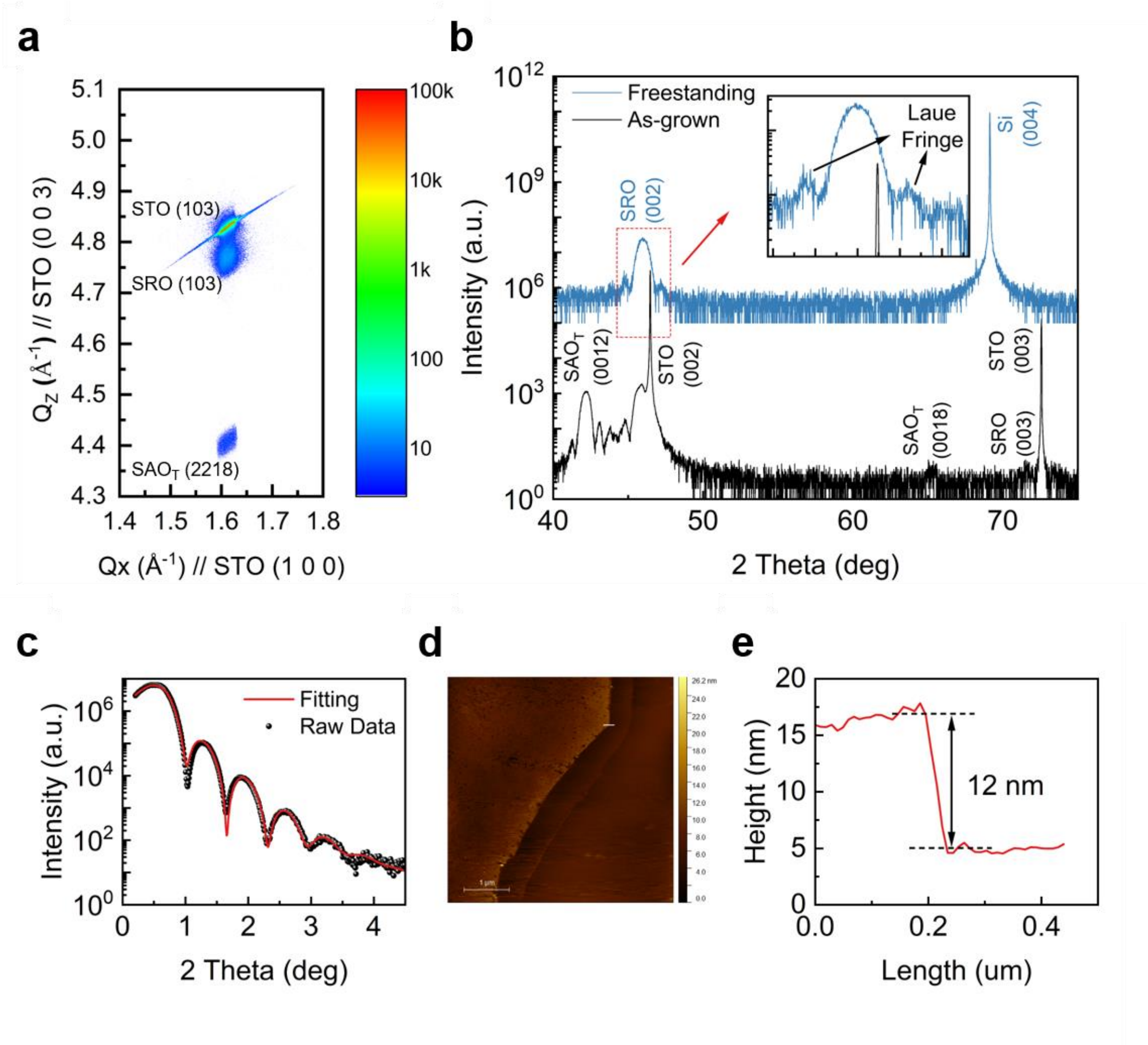

**Figure 3 | Structure of SRO epitaxial film and freestanding membrane. (a)** Reciprocal space mapping of the epitaxial film, showing that both SAOT and SRO remain coherently aligned with the STO substrate without detectable secondary phases. **(b)** XRD of the epitaxial film (black curve) and the SRO freestanding membrane integrated with Si (blue curve). **(c)** X-ray reflectivity of the SRO freestanding membrane on Si, with raw data (black curve) and a GenX fitting data (red curve). **(d)** AFM surface morphology of the SRO freestanding membrane integrated with Si. **(e)** Thickness of the SRO freestanding membrane determined along the white line in **d**.

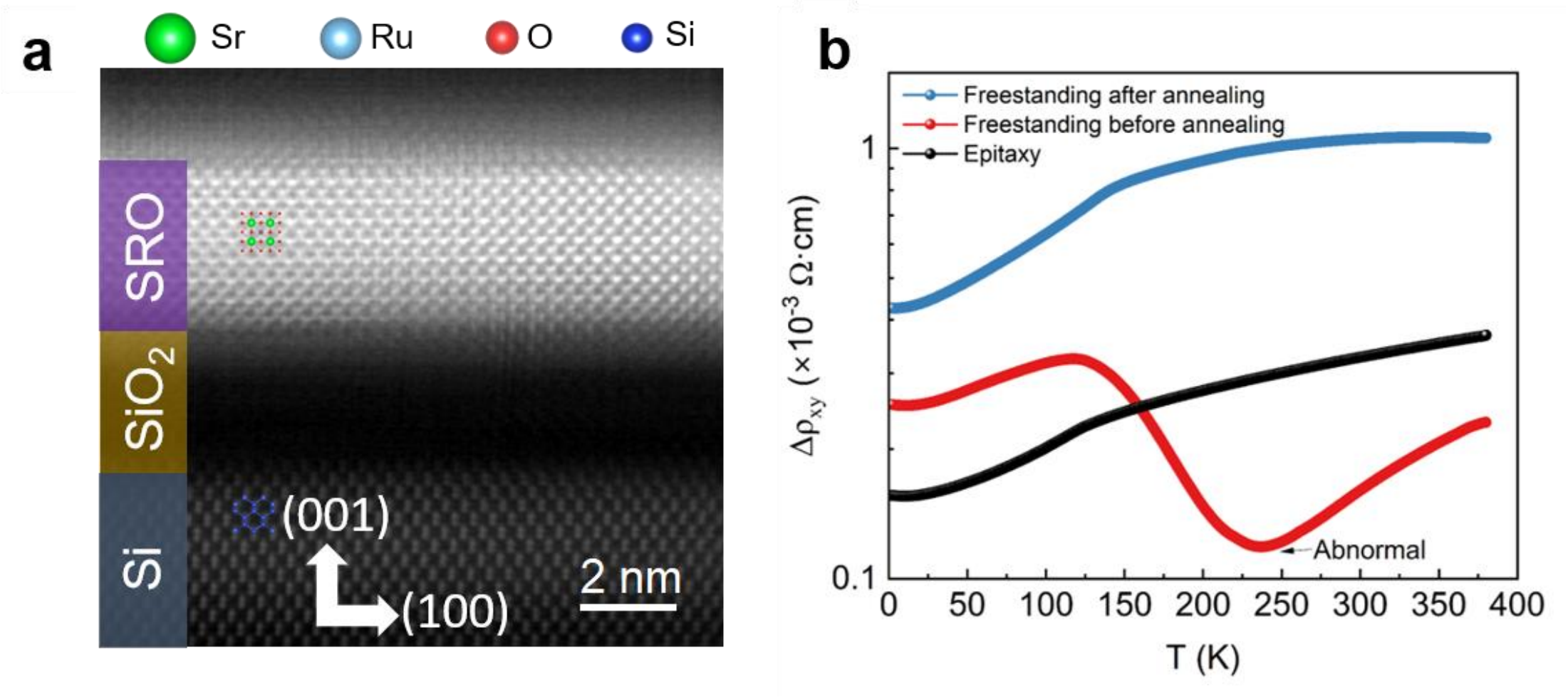

**Figure 4 | Ultrathin SRO freestanding membrane integrated with Si. (a)** High-angle annular dark-field STEM image of the SRO freestanding membrane with a thickness of 8-u.c.. Insets show the crystal structures of SRO (top) and the Si substrate (bottom). **(b)** Resistivity-temperature (R-T) curves of the epitaxial SRO film (black) and the freestanding membrane, with the red curve corresponding to the as-released state and the blue curve to the annealed membrane

Supplementary Materials for

# Ultrathin oxide freestanding membranes with large-scale continuity and structural perfection

Yuhao Hong[1, 2, 3, #, *], Yang Hu[1, 4, #], Jianyao Zheng[1], Minh D. Nguyen[1], Jelle R. H. Ruiters[1], Daniel M. Cunha[1], Iris C. G. van den Bosch[1], Edwin Dollekamp[3], Mark Huijben[1], Yulin Gan[2], Nini Pryds[3], Daesung Park[3], Christoph Baeumer[1], Qinghua Zhang[5], Guus Rijnders[1], Zhaoliang Liao[2, *], Gertjan Koster[1, *]

* Email: yuhho@dtu.dk, zliao@ustc.edu.cn, g.koster@utwente.nl

## 1. Sample growth

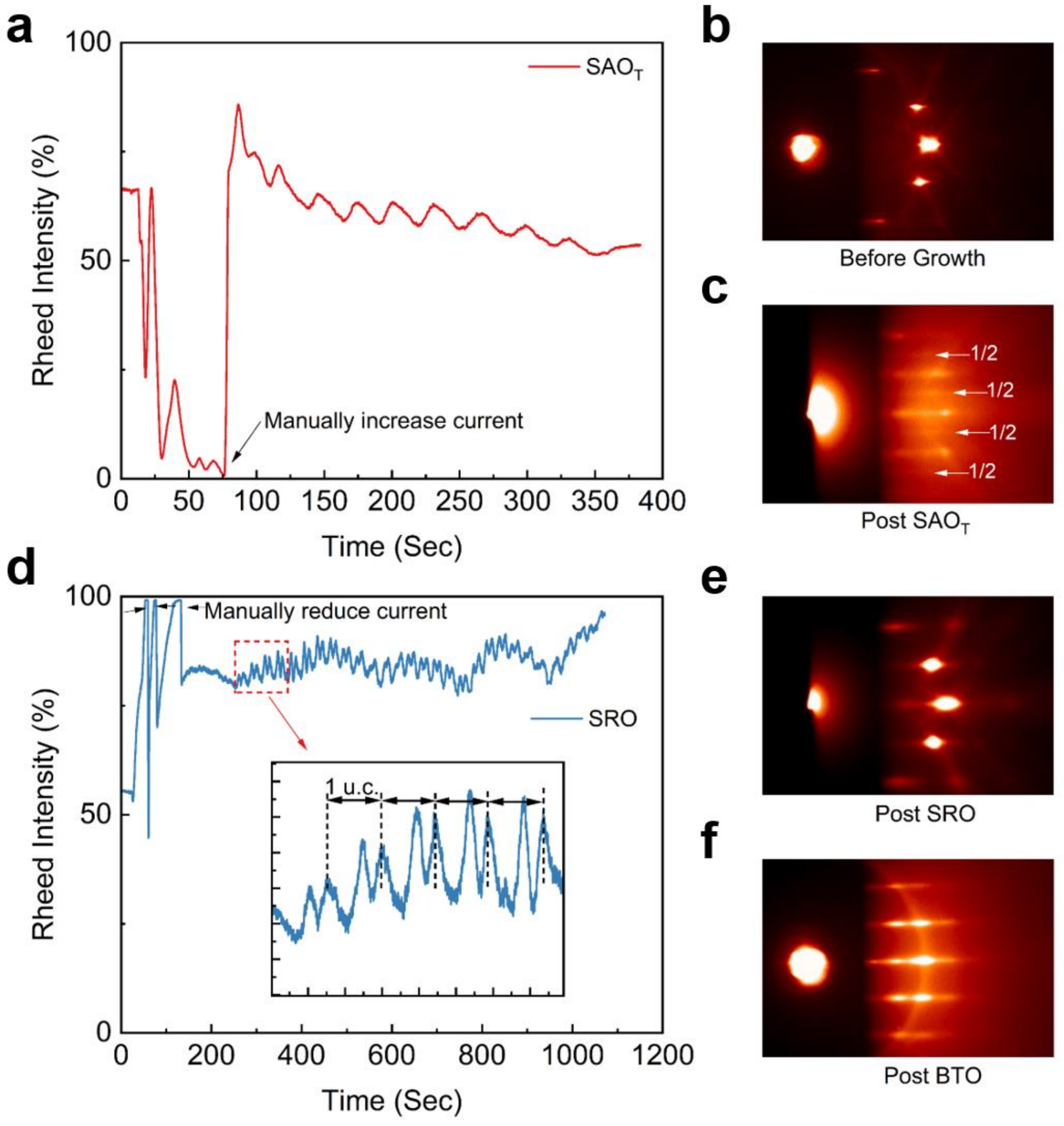

**Fig. S1 | RHEED characterization.** RHEED oscillations during the growth of **(a)** SAO$_T$ and **(d)** SRO. RHEED patterns **(b)** before growth and post growth of **(c)** SAO$_T$, **(e)** SRO, and **(f)** BTO.

## 2. Dissolution kinetics

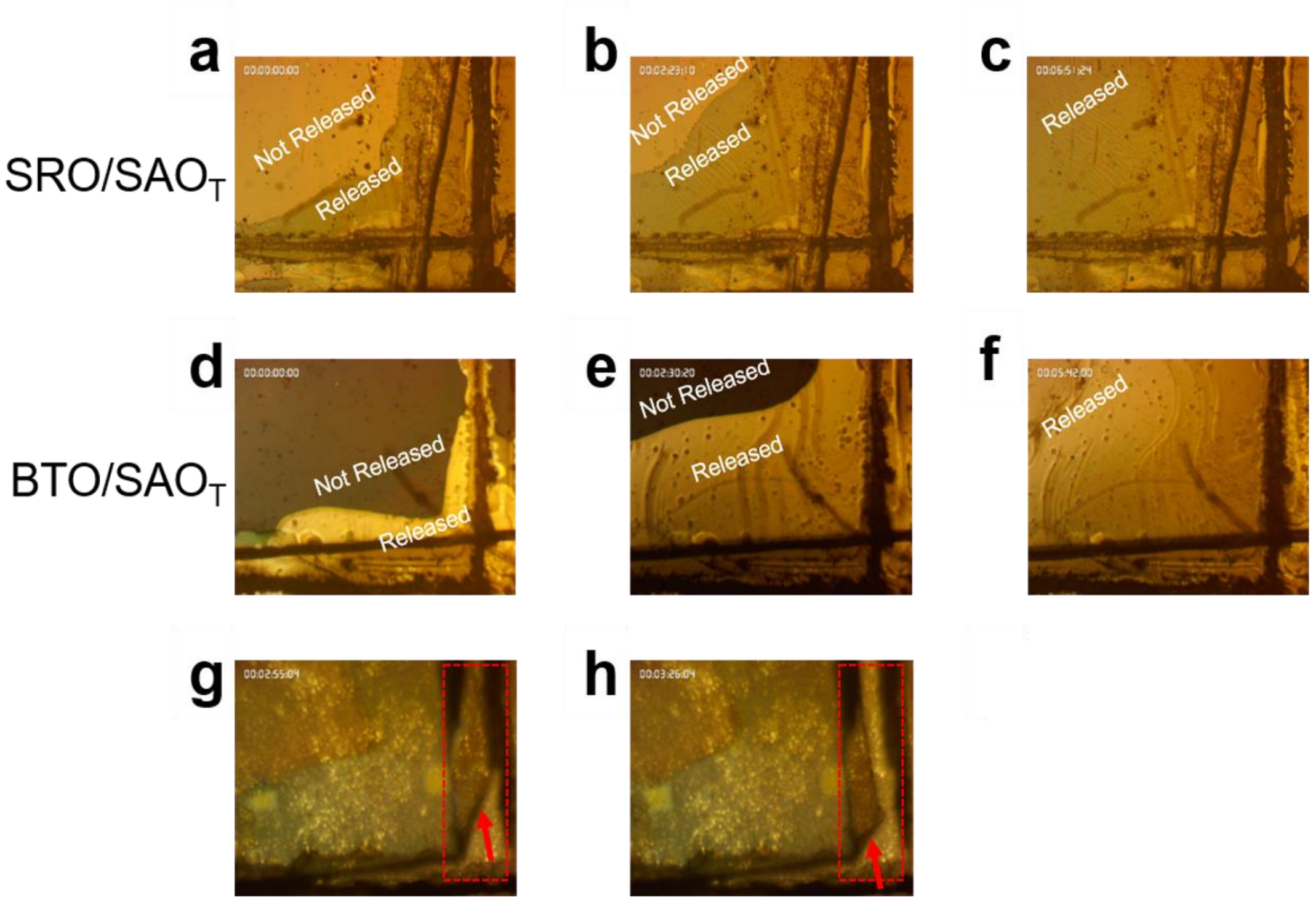

**Fig. S2 | Dissolution process of SAO$_T$ sacrificial layer. (a–c)** Dissolution process of the sacrificial layer in SRO/SAO$_T$ samples. **(d–f)** Dissolution process of the sacrificial layer in BTO/SAO$_T$ samples. Comparison of the uncovered PDMS region (Red dotted area) **(g)** before and **(g)** after sacrificial layer dissolution.

### 3. Optical morphology

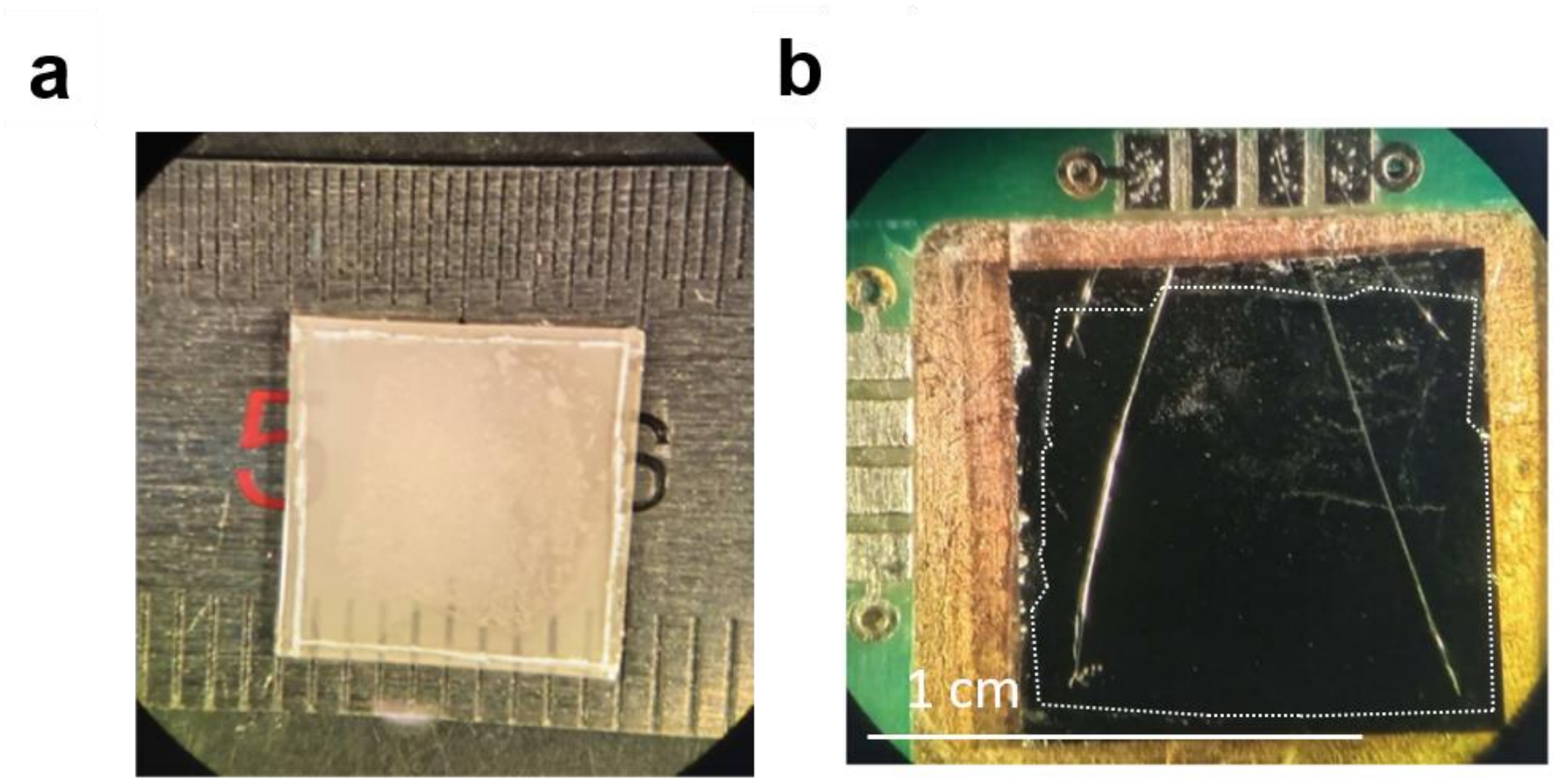

**Fig. S3 | Optical morphology. (a)** Optical morphology of the substrate surface after SRO transferred. **(b)** Configuration image of the freestanding SRO membrane integrated on Si during van der Pauw measurements.

## 4. Structural characterization of BTO freestanding membrane

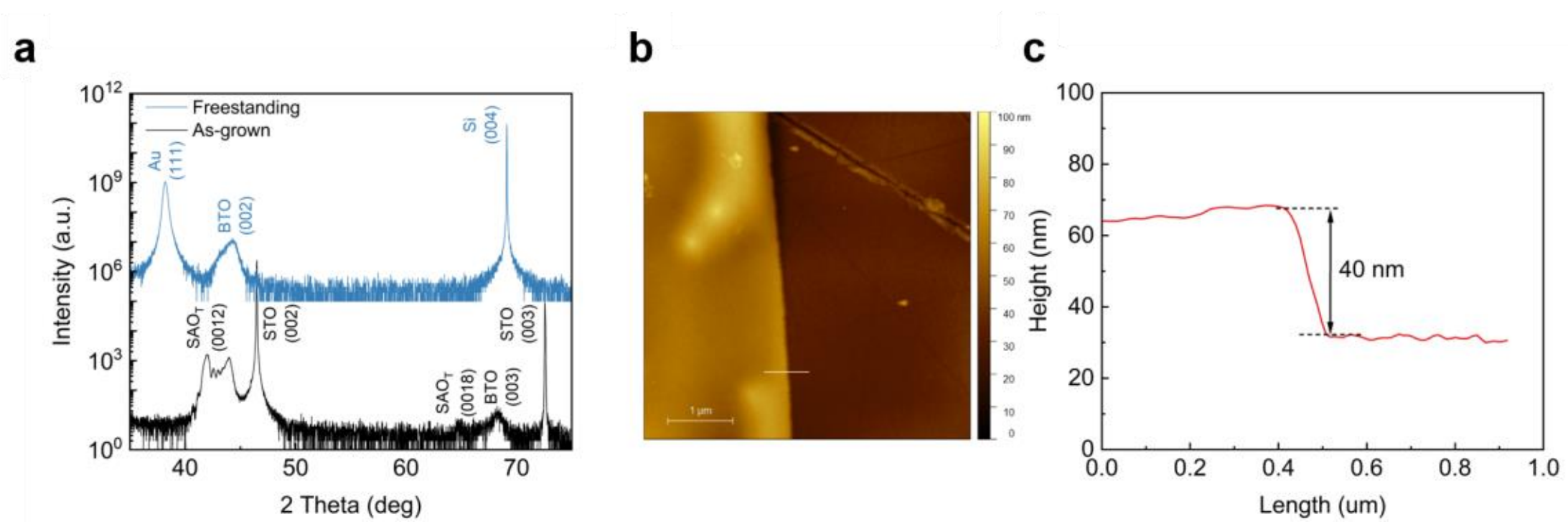

**Fig. S4 | Structural characterization of BTO freestanding membrane. (a)** XRD of the epitaxial film (black curve) and the BTO freestanding membrane integrated with Au/Si (blue curve). **(b)** AFM surface morphology of the BTO freestanding membrane integrated with Au/Si. **(e)** Thickness of the BTO freestanding membrane determined along the white line in **b**.

## 5. STEM images

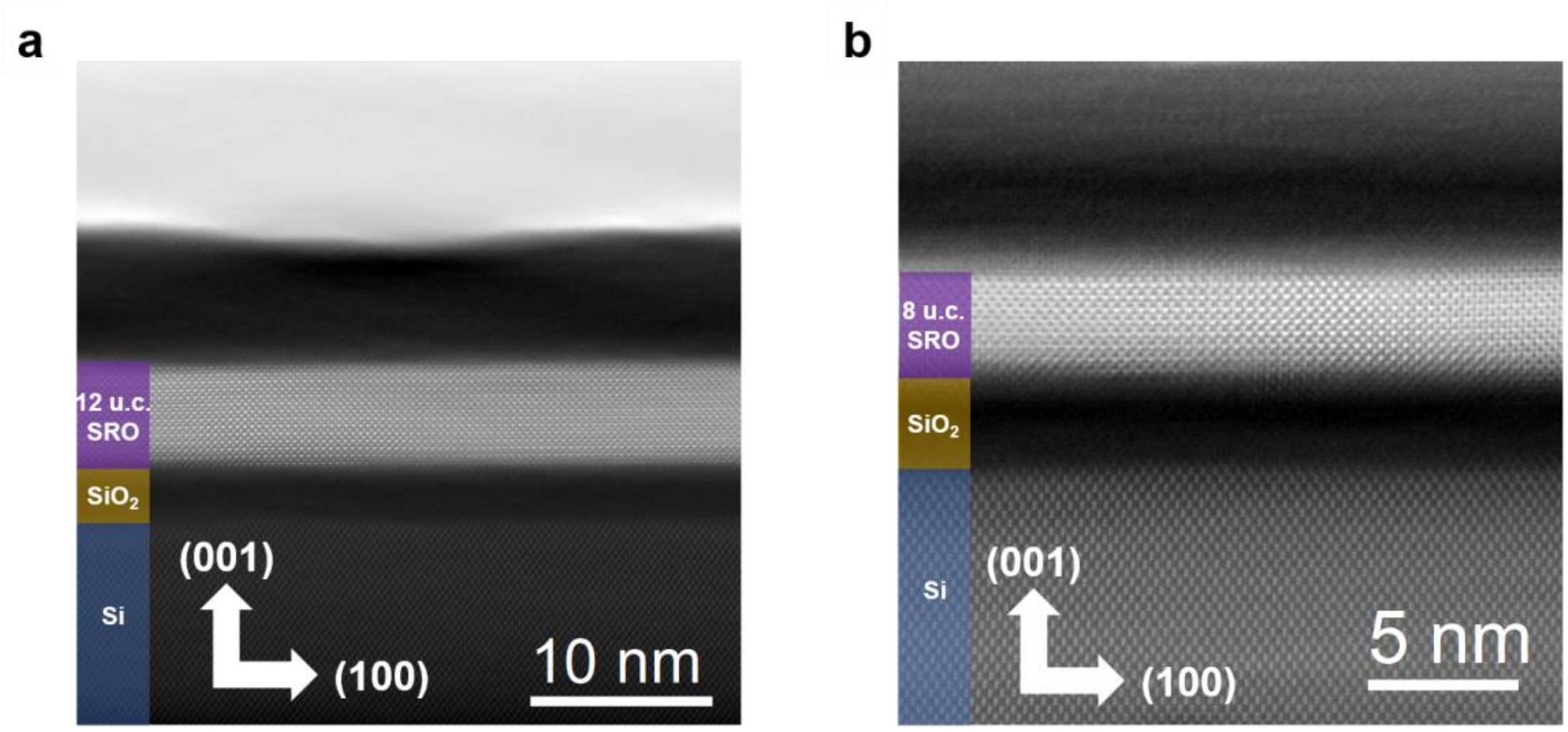

**Fig. S5 | STEM-HAADF image of ultrathin SRO freestanding membranes.** STEM-HAADF images of SRO freestanding membranes with thicknesses of **(a)** 12 u.c. and **(b)** 8 u.c..

## 6. Magnetoresistance of SRO freestanding membrane.

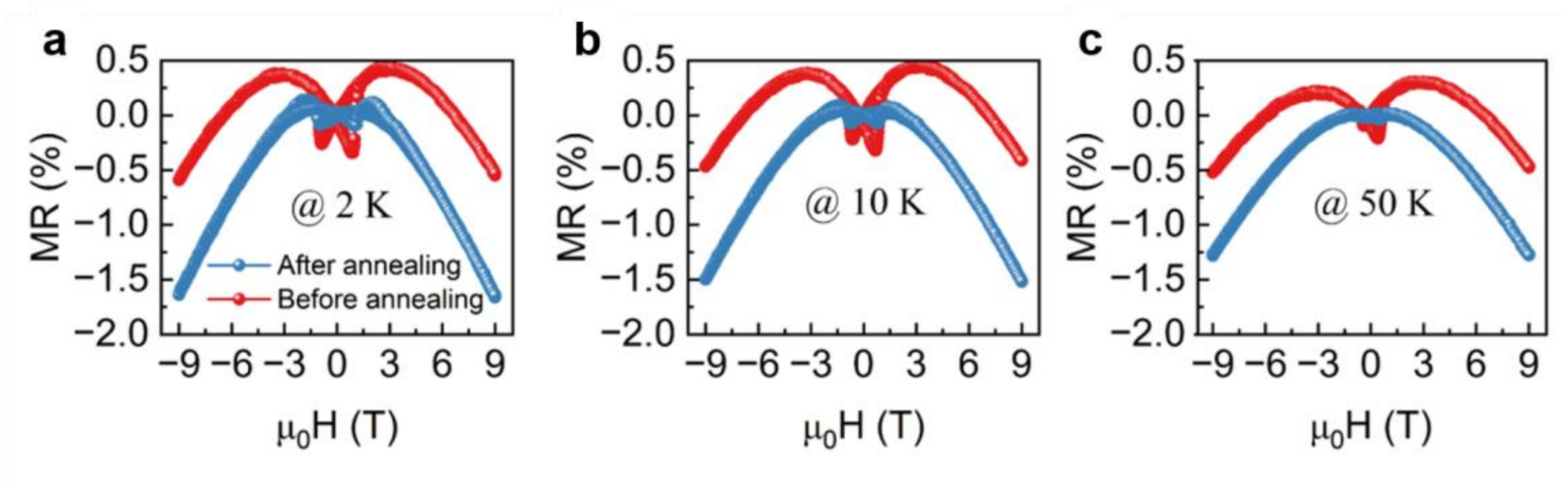

**Fig. S6 | Temperature-dependent magnetoresistance measurement of SRO independent membrane with thickness of 12 u.c.** Magnetoresistance of SRO freestanding membrane before (red spots) and after (blue spots) annealing at **(a)** 5K, **(c)** 10K and **(c)** 50K.

## 7. Oxygen vacancy distribution

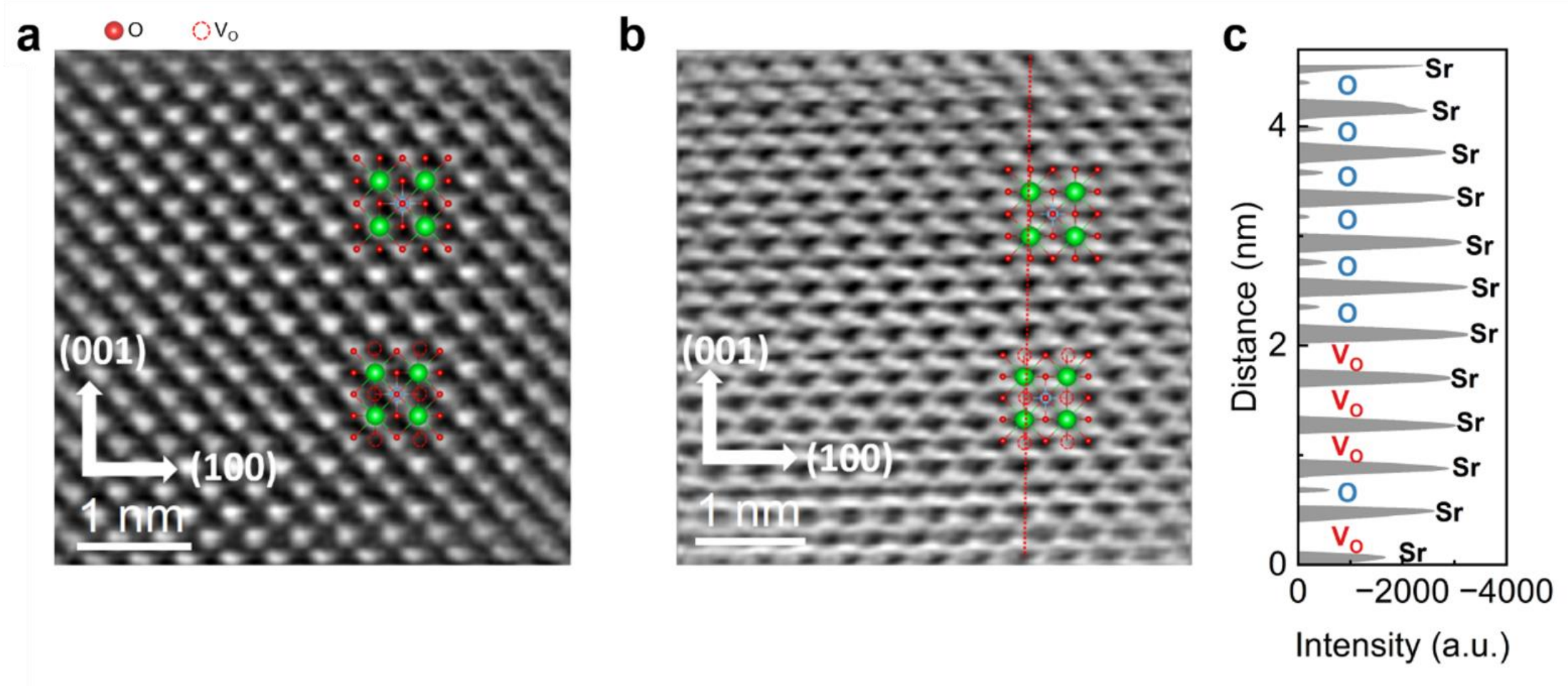

**Fig. S7 | Oxygen vacancies in ultrathin SRO freestanding membrane. (a)** STEM-HAADF and **(b)** STEM-ABF images of a 12-u.c.-thick SRO freestanding membrane, where filled circles mark oxygen atoms and open circles denote oxygen vacancies (inset). **(c)** Atomic distribution along the red dashed line indicated in **b.**

## 8. Ferroelectric characterization of BTO freestanding films

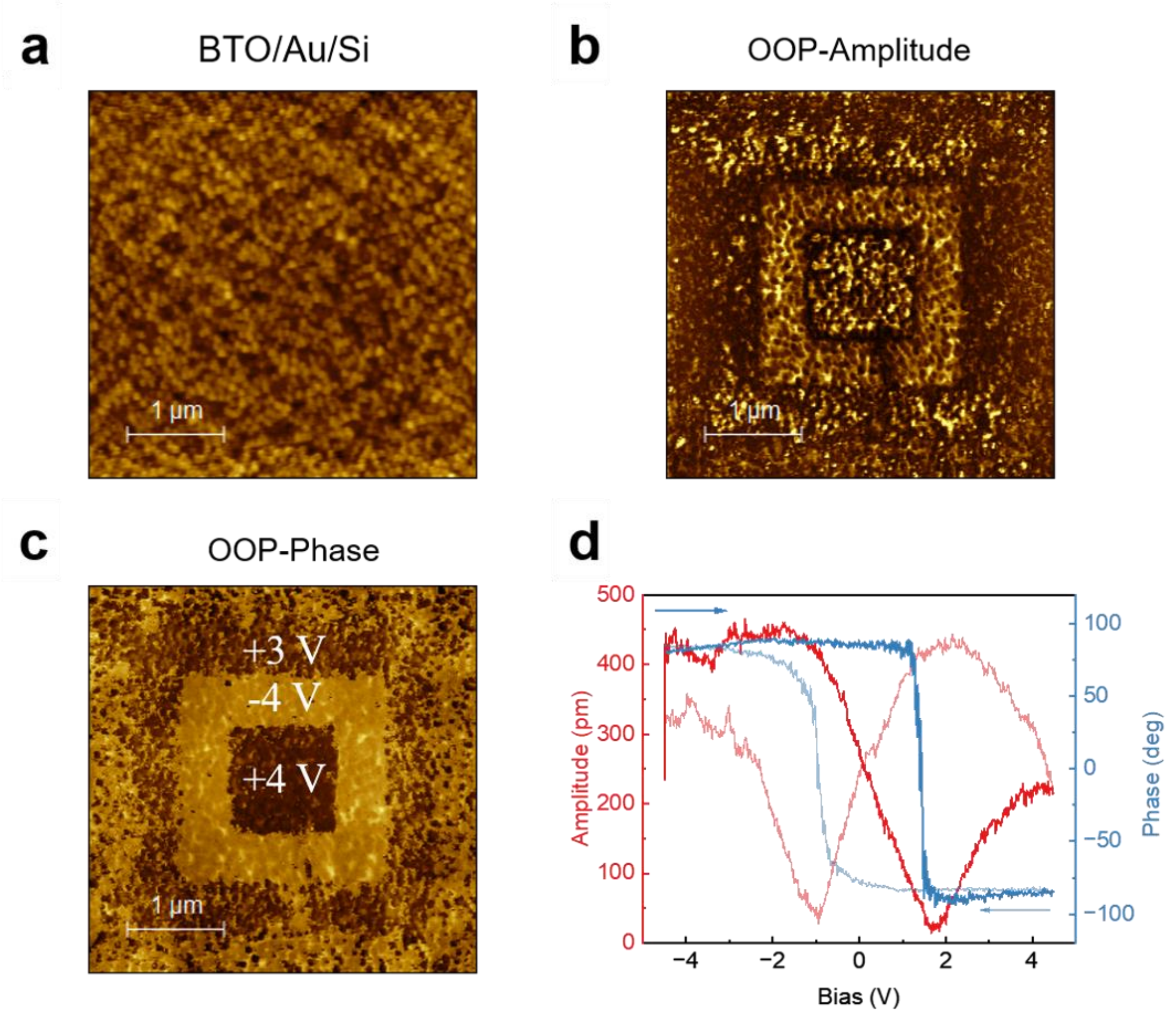

**Fig. S8 | Out-of-plane ferroelectric properties of the freestanding BTO membrane integrated with Au/Si. (a)** Surface morphology; **(b)** Out-of-plane amplitude; **(c)** Out-of-plane phase; **(d)** Ferroelectric hysteresis, where the red and blue curves correspond to amplitude and phase, respectively.

| *Layer* | *Thickness (nm)* | *Roughness (nm)* |
|---|---|---|
| $SrRuO_3$_ | 11.98 | 1.16 |
| $SiO_2$_ | 1.82 | 1.32 |
| Si Substrate | | 0.39 |

**Table S1 | Fitting parameters obtained from the XRR curve in Fig. 3c by GenX.**